\documentclass{ws-procs975x65}    
\usepackage{graphicx}% Include figure files
\usepackage{subfigure}
\usepackage{mathtools}
\usepackage{wrapfig}% Align table columns on decimal point% bold math
\newcommand{\apj}{ApJ}

\begin{document}
\title{Eccentric equatorial trajectories around a Kerr black hole as a QPO model for M82X-1}

\author{Prerna Rana$^{1, \dagger}$ and  A.\ Mangalam$^{2, \ddagger}$}

\address{$^1$Department of Astronomy and Astrophysics, Tata Institute of Fundamental Research, Homi Bhabha Road, Navy Nagar, Colaba, Mumbai 400005, India \\ $^2$ Indian Institute of Astrophysics, Sarjapur Road, 2nd Block Koramangala, Bangalore, 560034, India \\
E-mail: prerna.rana@tifr.res.in$^\dagger$, mangalam@iiap.res.in$^\ddagger$}

\begin{abstract}
We study the bound orbit conditions for equatorial and eccentric orbits around a Kerr black hole both in the parameter space ($E$, $L$, $a$) representing the energy, angular momentum of the test particle, and spin of the black hole, and also ($e$, $\mu$, $a$) space representing the eccentricity, inverse-latus rectum of the orbit, and spin. We apply these conditions and implement the relativistic precession (RP) model to M82X-1, which is an Intermediate-mass black hole (IMBH) system, where two high-frequency Quasi-Periodic Oscillations (HFQPOs) and a low-frequency QPO were simultaneously observed. Assuming that the QPO frequencies can also be generated by equatorial and eccentric trajectories, we calculate the probability distributions to infer $e$, $a$, and periastron distance, $r_p$, of the orbit giving rise to simultaneous QPOs. We find that an eccentric orbit solution is possible in the region between innermost stable circular orbit (ISCO) and the marginally bound circular orbit (MBCO) for $e=0.2768^{+0.0657}_{-0.0451}$,  $a=0.2897 \pm 0.0087$, and $r_p=4.6164^{+0.0694}_{-0.1259}$. 
\end{abstract}

\keywords{Classical black holes; Relativity and Gravitation; Infall, accretion and accretion disks; Kerr black hole; QPOs.}

\bodymatter

%%%%%%%%%%%%%%%%% now a standard article style for the most part

\section{Introduction}
Black holes have been discovered  as part of various systems in galaxies; as one of the components of the transient binary systems known as black hole X-ray binaries (BHXRB) with black hole mass in the range $5-30 M_{\odot}$, as intermediate mass black holes (IMBH) with mass range 300 to 600 $M_{\odot}$, and as galactic nuclei of mass $10^{6}-10^{10} M_{\odot}$ (Ref.~\refcite{NarayanMcClintock2013}). X-ray emission from such systems gives evidence of the presence of a compact object as X-rays originate very close to the black hole, and hence are important probes of the strong gravity regime. One of the important features discovered in the X-ray emission from these systems is quasi-periodic oscillations (QPOs), which are broad peaks in the Fourier power density spectrum. QPOs have been discovered with their frequencies ranging from mHz-kHz (Ref.~\refcite{BelloniStella2014}), in BHXRB, mHz to a few Hz in IMBH (Ref.~\refcite{Pasham2014}), and expected with the time scales of hours to days for AGN. The most interesting features among these are the high-frequency QPOs (HFQPOs) in the range 100-500Hz (Refs.~\refcite{Strohmayer2001a,Remillard2002}), detected in various BHXRB which may be used to calculate spin $a$ and mass $M_{\bullet}$ of the black hole when detected simultaneously. Some of these objects have also shown two simultaneous HFQPOs with their centroid frequencies having ratios $\sim$3:2 or 5:3 (Ref.~\refcite{BelloniStella2014}), indicating a resonance. There are also some cases, for example, GROJ1655-40 (Ref.~\refcite{Motta2014a}), and M82 X-1 (IMBH) (Ref.~\refcite{Pasham2014}), which are known to show the detection of three simultaneous QPOs, two HFQPOs and one low-frequency QPO (LFQPO). There are various existing models that attempt to explain the origin of QPOs but perhaps the most widely accepted is the relativistic precession (RP) model (Ref.~\refcite{Stella1999b}), which interprets LFQPO as the nodal precession frequency given by $\left( \nu_{\phi}\right.$-$\left.\nu_{\theta}\right) $, and the two HFQPOs as associated with the azimuthal frequency, $\nu_{\phi}$, and the periastron precession frequency $\left( \nu_{\phi}\right.$-$\left.\nu_{r}\right)$ of a particle orbiting around a black hole. RP model has been previously applied to the cases of BHXRB (Ref.~\refcite{Motta2014a}), assuming that the frequencies correspond to circular orbits of an equatorial system. 

In this article, we first present the necessary bound orbit conditions both in the ($E$, $L$, $a$) and in the ($e$, $\mu$, $a$) space of the corresponding orbit for the equatorial bound orbits. Using these conditions, we classify various bound orbits in both spaces. Then, by applying the RP model to the case of an IMBH M82X-1, we find that an eccentric equatorial orbit can also originate three simultaneous QPOs. We also discuss the possible region around the black hole giving rise to these QPOs.

\section{Bound orbit conditions for equatorial orbits}
In this section, we write the bound orbit conditions in ($E$, $L$) and ($e$, $\mu$) space for a particle orbiting a Kerr black hole of a given spin, $a=J/M_{\bullet}$. Throughout this article, we have scaled $E$, $L$, and $a$ parameters by mass of the black hole $M_{\bullet}$, and have used geometrical units ($G=c=1$).

\subsection{\underline{Dynamical parameter space $\{E, L, a \}$}}
\label{ELaspace}
\begin{figure}[ht!]
\begin{center}
\mbox{ \subfigure[]{
\includegraphics[width=5.5cm,height=3.5cm]{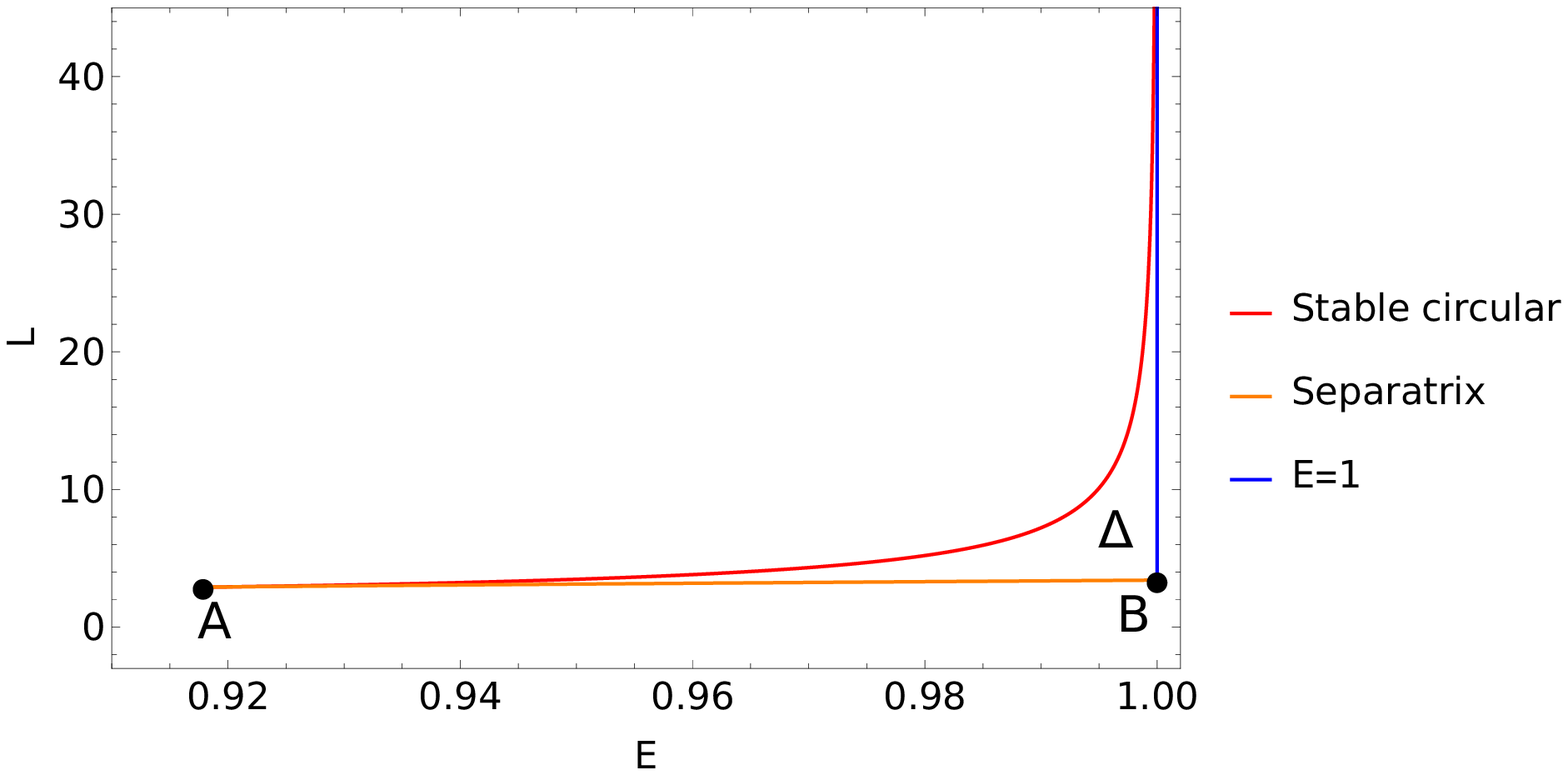} \label{deltacondition2a}}
\qquad
\subfigure[]{
\includegraphics[width=5.5cm,height=3.5cm]{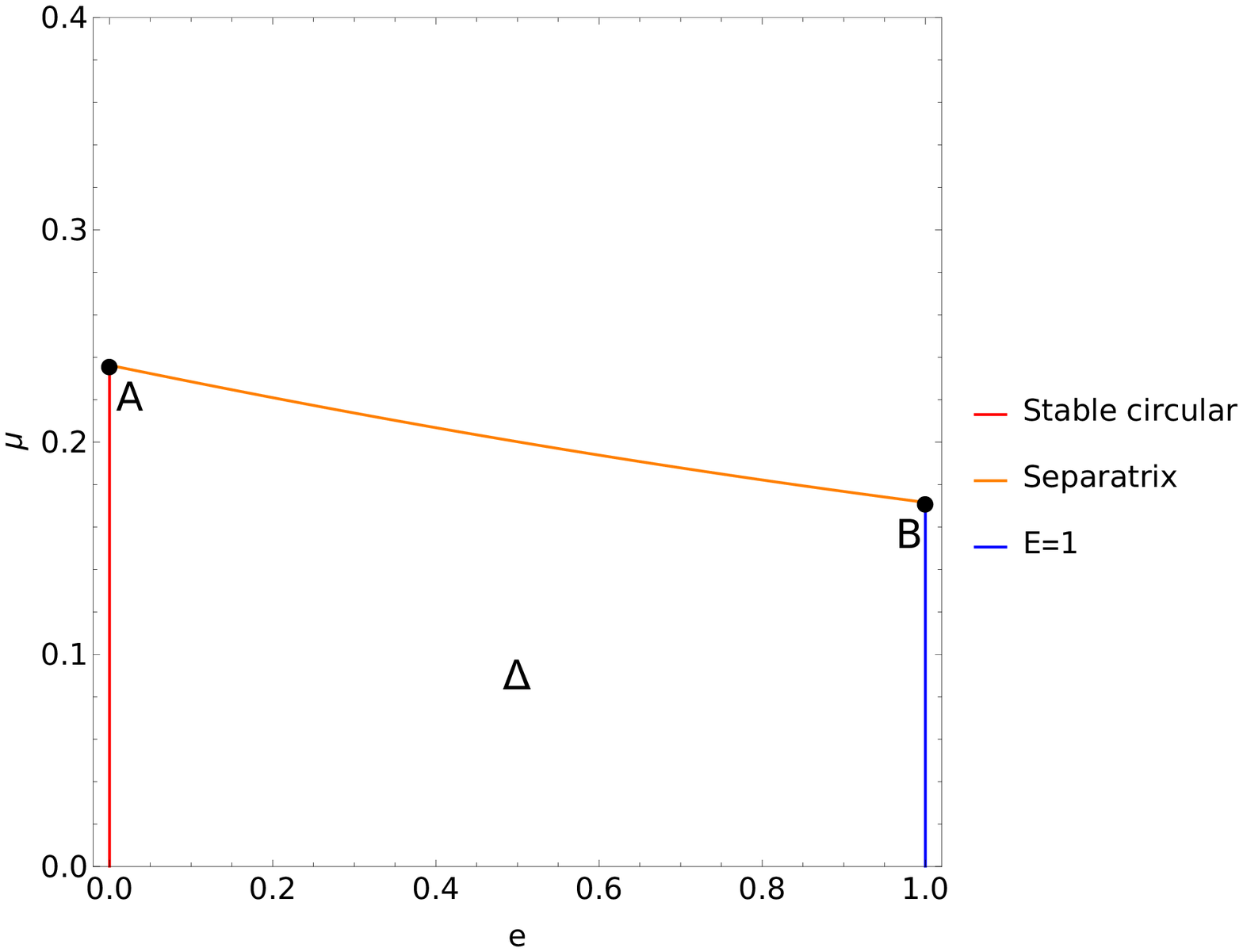}\label{deltacondition2b}}}
\end{center}
\caption{\label{deltacondition2} The bound orbit condition, $\Delta\leq0$, is shown as the $\Delta$ region (a) bounded by stable circular Eq. (\ref{stabcond}), separatrix orbits Eq. (\ref{sepcond}), and $E=1$ in the ($E$, $L$) plane, and (b) bounded by stable circular Eq. (\ref{e0}), separatrix orbits Eq. (\ref{factor1}), and $e=1$ in the ($e$, $\mu$) plane for $a=0.5$, where points A and B represents the innermost stable circular orbit (ISCO) and marginally bound circular orbit (MBCO) respectively.}
\end{figure}

The equation defining radial motion of a particle (unit mass, $m_0=1$) in the equatorial plane of a Kerr black hole having spin $a$ is given by (Ref.~\refcite{Carter1968})
\begin{equation}
\dfrac{\left( E^2 - 1 \right)}{2}=\dfrac{1}{2}\left( \dfrac{{\rm d}r}{{\rm d} \tau}\right)^2 -\dfrac{1}{r} \ + \ \dfrac{L^{2}-a^{2}\left( E^{2}-1\right)  }{2r^{2}}-\dfrac{\left( L-aE\right)^{2}}{r^{3}}, \label{radialeqn1}
\end{equation}
where $r$ is the radial distance from black hole and $\tau$ is the proper time. The radial kinetic energy vanishes at the turning points of the orbit, ${\rm d}r /{\rm d} \tau =0$, which gives
\begin{equation}
\left( 1-E^{2}\right)  r^{3} \ - \ 2r^{2} \ + \ \left( L^{2}-a^{2}\left( E^{2}-1\right)  \right)  \ r \ - \ 2(L- a E)^{2}=0.
\label{cubeqn}
\end{equation}
 We apply the Cardano's method (Ref.~\refcite{Cardano}) to obtain real roots of Eq. (\ref{cubeqn}) in terms of dynamical parameters $\{E, L, a \}$, which corresponds to the case when a bound orbit exists between the first two turning points. The three real roots of Eq. (\ref{cubeqn}) can be expressed as (Ref. \refcite{RM2019}) 
 \begin{subequations}
 \begin{eqnarray}
 r_{1}=&&r_{a}=\dfrac{2\left[  1+ \left[4-3\left( 1-E^{2}\right) \left( L^{2}-a^{2}\left( E^{2}-1\right) \right)  \right] ^{1/2} \cos\left( \dfrac{\varphi}{3}\right) \right] }{3\left( 1-E^{2}\right) }, \label{root1} \\
 r_{2}=&&r_{p}=\dfrac{2\left[  1+ \left[4-3\left( 1-E^{2}\right) \left( L^{2}-a^{2}\left( E^{2}-1\right) \right)  \right] ^{1/2} \cos\left( \dfrac{\varphi- 2\pi}{3}\right)  \right] }{3\left( 1-E^{2}\right) }, \label{root2}\\
 r_{3}=&&\dfrac{2\left[  1+ \left[4-3\left( 1-E^{2}\right) \left( L^{2}-a^{2}\left( E^{2}-1\right) \right)  \right] ^{1/2} \cos\left( \dfrac{\varphi+ 2\pi}{3}\right) \right] }{3\left( 1-E^{2}\right)},  \label{root3}
  \end{eqnarray}
where $\varphi$ is defined by
\begin{equation}
 \cos\varphi =\dfrac{\left[ 8-9\left( 1-E^{2}\right)\left( L^{2}-a^{2}\left( E^{2}-1\right) \right) + 27\left( 1-E^{2}\right)^{2} \left( L- a E\right) ^{2}\right] }{\left[ 4-3\left( 1-E^{2}\right) \left( L^{2}-a^{2}\left( E^{2}-1\right) \right) \right] ^{3/2}}; \label{cosphiEL}
\end{equation} 
where $r_{1}>r_{2}>r_{3}$, $r_{1}$ and $r_{2}$ are the apastron and periastron points of the eccentric orbit respectively. It follows that 
\begin{equation}
-1<\cos \varphi<+1 \quad \mathrm{holds \ for \ eccentric \ orbits}. \label{cosphiecc}\\
\end{equation}
\end{subequations}
As per the Cardano's method, the condition for three real roots is described by the discriminant of the cubic equation, which we obtain for Eq. \eqref{cubeqn}, given by (Ref.~\refcite{RM2019})
\begin{eqnarray}
\Delta=&&27\left( 1-E^{2}\right) ^{2}x^{4}-  \left( L^{2}-a^{2}\left( E^{2}-1\right) \right) ^{2} -18x^{2}\left( 1-E^{2}\right) \left( L^{2}-a^{2}\left( E^{2}-1\right) \right) \nonumber \\
&& + 16x^{2}  +  \left( 1-E^{2}\right)  \left( L^{2}-a^{2}\left( E^{2}-1\right) \right) ^{3},
\label{deltaEL}
\end{eqnarray}
where $x=L-aE$. Hence, the condition on $\{E, L, a \}$ to get three real and distinct roots of $r$, which is possible only when the orbit is an eccentric bound orbit, is given by 
\begin{eqnarray}
\Delta<0 \qquad \mathrm{and} \qquad 0<E<1. 
\label{ecc.cdn}
\end{eqnarray}
The case of two equal roots and one distinct real root is possible when $\Delta=0$,
which corresponds to the stable circular, and the separatrix orbit or the unstable circular orbit, which are classified using $\cos\varphi$ by the following conditions:
\begin{subequations}
\begin{eqnarray}
\mathrm{Stable \ circular \ orbit \ when} \ \cos \varphi =&&-1 \ \Rightarrow \  r_{1}=r_{2} \ \ \mathrm{and} \ \ 0<E<1, \label{stabcond} \\
\mathrm{Separatrix \ orbit \ when} \ \cos \varphi =&&+1 \ \Rightarrow \  r_{2}=r_{3} \ \ \mathrm{and} \ \ 0<E<1, \label{sepcond} \\
 \mathrm{Unstable \ circular \ orbit \ when} \ \cos \varphi =&&+1 \ \Rightarrow \ r_{2}=r_{3} \ \ \mathrm{and} \ \ E>1. \label{unstabcond}
\end{eqnarray}
\end{subequations}
In this way, we are able to classify the various bound orbits in the $\{E, L, a \}$ space. Fig. \ref{deltacondition2a} shows this classification as bound orbit region, we call it the $\Delta$ region, in the ($E$, $L$) plane which is bounded by the curves representing the stable circular orbits Eq. (\ref{stabcond}), separatrix orbits Eq. (\ref{sepcond}), and $E=1$. The details of results presented in this section are provided in Ref.~\refcite{RM2019}. 

\subsection{\underline{Conic parameter space $\{e, \mu, a \}$}}
\label{emuaspace}
Next, we extend and translate the bound orbit conditions to the conic parameter space which is useful for the geometric study of the bound trajectories. The eccentricity and inverse-latus rectum are defined as
\begin{equation}
e=\dfrac{r_{a}-r_{p}}{r_{a}+r_{p}},  \ \
\mu=\dfrac{r_{a}+r_{p}}{2r_{a}r_{p}}.
\end{equation}
Using the transformation relations between ($E$, $L$) and ($e$, $\mu$) (Refs.~\refcite{RMCQG2019,RM1arxiv2019}), we can express $\Delta$, Eq. (\ref{deltaEL}), in terms of $\{e$, $\mu$, $a\}$ as (Ref.~\refcite{RM2019})
 \begin{eqnarray}
 \Delta =&&\dfrac{-e^{2}}{\mu^{2}}\left[ 1-x^{2}\mu^{2}\left( 1+ e\right) \left(3-e \right) \right] ^{2} \left[1-x^{2}\mu^{2}\left( 1- e\right) \left(3+e \right) \right] ^{2}= \dfrac{-e^{2}}{\mu^{2}} \Delta_{1}^{2} \Delta_{2}^{2}. \nonumber \\
 \label{deltaemu}
 \end{eqnarray}
It is seen that $\Delta\leq 0$ is valid for all the real values of $x$, and $\Delta=0$ if any one of the following conditions is satisfied:
\begin{subequations}
\begin{eqnarray}
e=0, 
\label{e0}\\
\Delta_{1}=0\Rightarrow \mu^{2}x^{2}\left( 3-e\right) \left( 1+e\right)  = 1, 
\label{factor1}\\
\Delta_{2}=0\Rightarrow \mu^{2}x^{2}\left( 3+e\right) \left( 1-e\right)  = 1, 
\label{factor2}
\end{eqnarray}
\end{subequations}
where Eq. (\ref{e0}) corresponds to the stable circular orbits, Eq. (\ref{factor1}) corresponds to the separatrix orbits and they together with $e=1$ define the boundaries of the $\Delta$ region in the ($e$, $\mu$) plane; see Fig. \ref{deltacondition2b}. 
Hence, the condition in $\{e, \mu,  a \}$ space for an eccentric bound orbit is given by $r_{3}<r_{p}$, which corresponds to (Ref.~\refcite{RM2019})
\begin{equation}
\mu^{2}x^{2}\left( 3-e\right) \left( 1+e\right)  \leq 1,
\label{odrrootcdn}
\end{equation}
and is the operative condition for eccentric bound orbits representing the ordering of roots, $r_{1}>r_{2}>r_{3}$.
The expression of $\cos \varphi$, Eq. \eqref{cosphiEL}, in the $\{e$, $\mu$, $a\}$ space is given by (Ref.~\refcite{RM2019})
\begin{equation}
\cos\varphi= \dfrac{\splitfrac{\left[1- 3 x^{2}\mu^{2}\left( 1- e^{2}\right)  \right]\cdot \left[ -1 + 3 e + 3 x^{2} \mu^{2}\left( 1- e -e^{2}+ e^{3}\right)\right] \cdot}{\left[1+ 3e - 3 x^{2} \mu^{2}\left( 1+ e - e^{2}- e^{3}\right)  \right]  } }{\left\lbrace  4-3 \left(1-e^2 \right) \left[ 1-x^2 \mu^2 \left(1-e^2 \right)\right] \left[1+ \mu^2 x^2 \left(3+e^2 \right) \right] \right\rbrace ^{3/2} } .
\label{cosphiemua}
\end{equation}

\section{Parameter estimation from RP model for QPOs}
\label{RPQPOapp}
\begin{table}
\tbl{Fundamental frequencies of}{ equatorial eccentrc orbits, Refs.~\refcite{RMCQG2019,RM1arxiv2019,RMQPO2019}. }
\begin{center}
\scalebox{0.656}{
\begin{tabular}{|c|c|}
\hline
& \\
$\nu_{\phi}\left(e, \mu, a \right)$ & $\dfrac{c^{3}\cdot \left\lbrace  a_{1} \mathrm{EllipticPi}\left[ -{p_{2}}^{2}, \pi /2, m^{2}\right] + b_{1} \mathrm{EllipticPi}\left[ -{p_{3}}^{2},\pi /2, m^{2}\right]  \right\rbrace }{2 \pi G M\left\lbrace \splitfrac{ a_2 \left[ \dfrac{{p_{1}}^{2}\mathrm{EllipticE}\left[ \pi /2, m^{2}\right] }{2\left(1+{p_{1}}^{2} \right) \left( m^{2} + {p_{1}}^{2}\right) } -\dfrac{\mathrm{EllipticF}\left[ \pi /2, m^{2}\right] }{2\left(1+{p_{1}}^{2} \right) } \right] + c_2 \mathrm{EllipticPi} \left[ -{p_{2}}^{2} , \pi /2 , m^{2}\right] }{+ \mathrm{EllipticPi}\left[ -{p_{1}}^{2} , \pi /2 , m^{2}\right]  \left\lbrace a_2  \dfrac{\left[ {p_{1}}^{4} +2{p_{1}}^{2} \left( 1+m^{2}\right) + 3m^{2}\right] }{2\left( 1+{p_{1}}^{2}\right) \left( m^{2} +{p_{1}}^{2}\right)} +b_2\right\rbrace + d_2 \mathrm{EllipticPi}\left[ -{p_{3}}^{2} , \pi /2 , m^{2}\right] } \right\rbrace  }$,\\
& \\
\hline
& \\
$\nu_{r}\left(e, \mu, a \right)$ & $\dfrac{c^{3}}{2 G M\left\lbrace \splitfrac{ a_2 \left[ \dfrac{{p_{1}}^{2}\mathrm{EllipticE}\left[ \pi /2, m^{2}\right] }{2\left(1+{p_{1}}^{2} \right) \left( m^{2} + {p_{1}}^{2}\right) } -\dfrac{\mathrm{EllipticF}\left[ \pi /2, m^{2}\right] }{2\left(1+{p_{1}}^{2} \right) } \right] + c_2 \mathrm{EllipticPi} \left[ -{p_{2}}^{2} , \pi /2 , m^{2}\right] }{+ \mathrm{EllipticPi}\left[ -{p_{1}}^{2} , \pi /2 , m^{2}\right]  \left\lbrace a_2  \dfrac{\left[ {p_{1}}^{4} +2{p_{1}}^{2} \left( 1+m^{2}\right) + 3m^{2}\right] }{2\left( 1+{p_{1}}^{2}\right) \left( m^{2} +{p_{1}}^{2}\right)} +b_2\right\rbrace + d_2 \mathrm{EllipticPi}\left[ -{p_{3}}^{2} , \pi /2 , m^{2}\right] } \right\rbrace   } $ \\
& \\
 \hline
 &\\
 $\nu_{\theta}\left(e, \mu, a \right)$ & $\dfrac{2 \nu_{r}a\sqrt{1-E^2}z_{+} \mu^{1/2} \mathrm{EllipticF}\left[ \dfrac{\pi}{2}, m^{2}\right] }{\pi \left[ 1- \mu^2 x^2 \left( 3 -e^2 -2e\right)  \right] ^{1/2}}$\\
 & \\
 \hline
 & \\
 & $m^2=\dfrac{4 \mu^2 e x^2}{\left[ 1- \mu^2 x^2 \left( 3 -e^2 -2e \right) \right]}$,  $ \ {p_{1}}^{2}= \dfrac{2e}{1-e}$, $ \ {p_{2}}^{2}=\dfrac{2e a^2 \mu}{\left( a^2 \mu - a^2 \mu e -r_{+}\right) }$, $ \ {p_{3}}^{2}=\dfrac{2e a^2 \mu}{\left( a^2 \mu - a^2 \mu e -r_{-}\right) }$, $\ {z_{+}}^2=\dfrac{\left[ a^2 \left( 1- E^2\right) + L^2 \right] }{a^2 \left(1-E^2 \right) }$,\\
 Constants & $a_1=\dfrac{\mu^{1/2}\left[ L a^2 -2 x r_{+}\right] }{\sqrt{1-a^2}\left(a^2 \mu- a^2 \mu e - r_{+} \right) \sqrt{1-\mu^2 x^2 \left(3 -e^2 -2 e \right) } }$, $b_1=\dfrac{\mu^{1/2}\left[ -L a^2 +2 x r_{-}\right] }{\sqrt{1-a^2}\left(a^2 \mu- a^2 \mu e - r_{-} \right) \sqrt{1-\mu^2 x^2 \left(3 -e^2 -2 e \right) } }$, 
\\
 &  $a_2= \dfrac{2  E}{\mu^{3/2}\left( 1-e\right)^2 \sqrt{1-\mu^2 x^2 \left(3 -e^2 -2 e \right) }}, \ \ b_2= \dfrac{4E}{\mu^{1/2}\left( 1-e\right) \sqrt{1-\mu^2 x^2 \left(3 -e^2 -2 e \right) } }$, \\
 &$c_2= \dfrac{2a^2 \mu^{1/2}\left(-La + 2 E r_{-}\right)}{r_{-}\sqrt{\left[ 1-\mu^2 x^2 \left(3 -e^2 -2 e \right)\right]  } \sqrt{1-a^2} \left( a^2 \mu -a^2 \mu e -r_{+}\right) }, \ \
d_2= \dfrac{2a \mu^{1/2}\left(-2 L r_{-}\sqrt{1-a^2}-2E r_{-}a +La^2\right)}{r_{-}\sqrt{\left[ 1-\mu^2 x^2 \left(3 -e^2 -2 e \right) \right] } \sqrt{1-a^2} \left( a^2 \mu -a^2 \mu e -r_{-}\right)}.$ \\
& \\
 \hline
\end{tabular}
\label{freqformulaeKerr}
}
\end{center}
\end{table}

Now, we take a more generalized approach to the RP model by considering $e\neq0$. We consider that three simultaneous QPOs can also be originated by a self-emitting blob orbiting in an equatorial eccentric trajectory around a Kerr black hole. Then, we calculate the parameters ($e$, $r_{p}$) of the orbit and spin $a$ of the black hole using the fundamental frequency formulae for equatorial orbits derived in Refs.~(\refcite{RMCQG2019,RM1arxiv2019}), and presented in Table \ref{freqformulaeKerr}. We search for the parameters in the allowed $\Delta$ region (shown in Fig. \ref{deltacondition2}) and defined by the bound orbit condition, Eq. \eqref{odrrootcdn}. We apply this method to the case of an IMBH M82X-1, which has shown three simultaneous QPOs at $\nu_1 =5.07\pm0.06$Hz, $\nu_2 = 3.32\pm0.06$Hz, and $\nu_3 =$204.8$\pm$6.3mHz (Ref.~\refcite{Pasham2014,Pasham2013ApJb}) to find $\{ e, r_p, a\}$. We fix the mass of black hole to be $M_{\bullet}=428M_{\odot}$ given in Ref.~\refcite{Pasham2014}, and simultaneously solve the equations $\nu_{\phi}=\nu_{1}$, $\nu_{\phi}-\nu_{r}=\nu_{2}$, and $\nu_{\phi}-\nu_{\theta}=\nu_{3}$ to find the exact solutions for the RP model given by $\{ e_0, {r_p}_0, a_0\}$. We assume the QPO frequencies as Gaussian distributed with their mean values at $\nu_{1}$, $\nu_{2}$ and $\nu_{3}$. We estimate the 1$\sigma$ errors in the solution $\{ e_0, {r_p}_0, a_0\}$ using the Jacobian of the transformation to $\{ \nu_1 , \nu_2 , \nu_3\}$ space whose formulae are given in Table \ref{freqformulaeKerr}, and find the corresponding normalized joint probability density distribution in the ($e$, $r_{p}$, $a$) space. As the probability density $P\left( e, r_p, a\right) $ is three dimensional, we collapse the profile of the probability density in two dimensions and fit a Gaussian function to find the mean values, $\{e_0, r_{p0}, a_0\}$, and variances, $\{\sigma_e, \sigma_{r_p}, \sigma_a \}$ in other dimension.

\begin{wrapfigure}{r}{0.48\textwidth}
  \vspace{1.5cm}
  \begin{center}
    \includegraphics[width=0.55\textwidth]{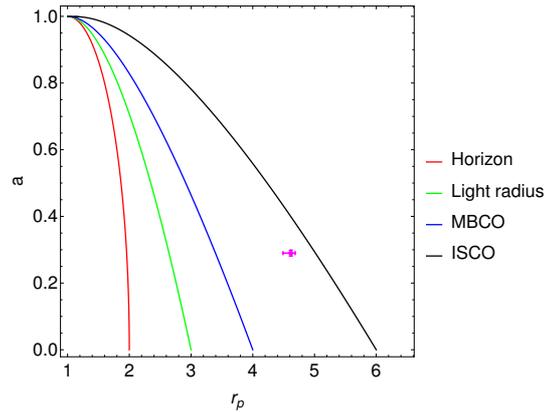}
  \end{center}
  \vspace{0pt}
  \caption{\label{M82X1}An equatorial eccentric orbit solution (the point in magenta color) shown in the ($r_p$, $a$) plane for the simultaneous QPOs discovered in M82X-1. The solution exists in the region between ISCO and MBCO.}
  \vspace{0pt}
\end{wrapfigure}

Using this procedure, we found the exact solution at $e=0.2768^{+0.0657}_{-0.0451}$,  $a=0.2897 \pm 0.0087$, and $r_p=4.6164^{+0.0694}_{-0.1259}$ for M82X-1. We see that the probability density naturally peaks at non-zero eccentricity suggesting that the most probable solution might not be restricted to circular orbits as was previously assumed in Ref.~\refcite{Motta2014a}. Fig. \ref{M82X1} shows this solution in the ($r_p$, $a$) plane suggesting an eccentric orbit having its periastron point in the region between ISCO and MBCO is the most probable orbit for the generation of three simultaneous QPOs. We have implemented a similar approach to various BHXRB to include non-equatorial and eccentric trajectories in Ref.~\refcite{RMQPO2019}.

\section{Summary and discussion}
\label{summary}
 We discussed the bound orbit conditions for the eccentric equatorial orbits in the $\{E, L, a\}$, and $\{ e, \mu, a\}$ spaces. We applied it to the specific case of M82X-1, where three simultaneous QPOs were discovered (Ref.~\refcite{Pasham2014,Pasham2013ApJb}), to find the parameters $e=0.2768^{+0.0657}_{-0.0451}$,  $a=0.2897 \pm 0.0087$, and $r_p=4.6164^{+0.0694}_{-0.1259}$ for the orbit generating these QPOs using the RP model. We find that the eccentric orbit solutions are possible in the region between ISCO and MBCO, as shown in Fig. \ref{M82X1}. Hence, by assuming that the accretion disk ends near ISCO, we conclude that the blobs that originate near ISCO and follow equatorial and eccentric trajectories in this region that produce HFQPOs. We also discuss the non-equatorial eccentric trajectories as possible solutions for QPOs assuming the RP model in Ref.~\refcite{RMQPO2019}. 
 
 We acknowledge the support from the SERB project CRG 2018/003415.

\vspace{0.1cm}
%\tiny

\end{document}